# Resilient Big Data Monetization

Rossi Kamal, Choong Seon Hong
Department of Computer Engineering
Kyung Hee University, South Korea

*Abstract*—*Big Data Monetization* is resembled by the domination of service provider's measurement tools over Smart-device user's common-interests (e.g. environmental information, emo-tion, etc.). However, adaptation to usage-dynamics necessitates stronger binding between common-interests and measurement-tools. Hence, *Resilient Big Data monetization* is devised as *k*-dominance and *m*-connectivity problems, such that common-interests are connected by *k*-ways to measurement tools, which are tied within each other in *m*-ways. Consequently, a greedy approximation algorithm *Plutus* (i.e resembling Greek god of wealth) is proposed, which isolates measurement tools to ac-quire domination over common-interests, establishes synergy from common-interests to measurement tools and then acquires divergence and sustains it within measurement tools. Hence, *Plutus* lays out the theoretical foundation of *Resilient fact-finding*, which is characterized by being decomposed into four fact-finding properties, namely, maximal-independence, influence, *k*-connectivity and *m*-dominance, respectively. Moreover, *Resilient Big Data Monetization* is a NP-hard problem, which is justified as amenable to greedy solution *Plutus*.

*Index Terms*—Big Data Monetization, Resilience, Service Man-agement.

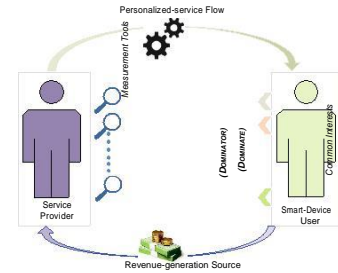

Fig. 1. Dominator, Dominate in *Resilient Big Data Monetization*

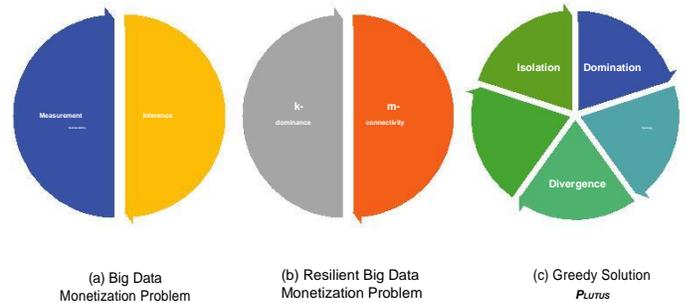

Fig. 2. (a) Big Data Monetization,(b) Resilient Big Data Monetization and Greedy Solution *Plutus*

## I. INTRODUCTION

### A. Motivation

The proliferation of personalized services inspires numerous service providers to monetize from Big Data or monitoring information of Smart-device-usage [1][2][3][4][5]. Hence, quantitative (e.g download-count, user-rating or usage-duration, etc) or qualitative (e.g. head-movement, taking tendency), application specific (e.g Ad-skipping behavior in Advertisement, video quality/buffering time in media content playing) measurement tools are often best-practiced nowadays[1][6][7]. Consequently, common-interests (e.g emotion, and environmental information) are often inferred through obtrusive or unobtrusive way to utilize in personalized shopping, search, well-being monitoring[2][8][9], traffic signal detection, consumer-appliances[3][10], air-pollution detection or even earthquake/nuclear threat detection[4][11][12]. In this context, Big Data monetization schemes are frequently devised by service providers to inspire prescribed campaign or strategy, truthful participation, lowest user-defined participation cost, consumer care, psychological insight, social strategy and robust participation[5]-[13].

### B. Resilient Big Data Monetization Problem

Hence, *Big Data Monetization* (Fig. 1) (Fig. 2(a)) is envisioned as being decomposed into two sub-problems, namely *Measurement* and *Inference*, so that service providers ac-quire dominance on measurement tools in inferring common-interests of Smart-device users. Therefore, it is formulated as *dominating set* problem (Theorem. 1), resembled by dominator (e.g. measurement-tools) and dominate (e.g common-interests), such that measurement-tools influence over common-interests.

Figuring out major measurement tools, in the midsts of qualitative, quantitative or innovative options, is a funda-mental requirement[14]. So, continuoal efforts[15][16] are required to acquire influence of measurement tools to adapt to usage-dynamics[14] or psychological insights (e.g expec-tation and purpose)[17] of users. In this context, some-times binding between common-interests and measurement tools becomes a necessity, especially in correlating activity-context, usage-frequency-social behavior, real-time feedback-worker relation[18][19][20]. Moreover, sometimes grouping among measurement assists to adapt, when observed in-formation are diversified, such as smartphone-traces (e.g. from waiting/on-the-bus people), ambient signs or even GPS face-to-face interaction on different time, within or outside vicinity[21][22]. Therefore, strengthening this binding among measurement tools, such as GPS-trace-verbal/proximity inter-action or even acoustic (e.g. ringtone and acoustic sound) and visual (e.g light intensity and view similarity) assists resilience to prevail[23][24][25][26].

In a nutshell, resilience in *Big Data Monetization*(Fig. 2(b)) necessitates isolation of influential measurement tools, which gradually should acquire sustainable domination over common-interests, who are also expected to be in binding with measurement tools. Hence, *Resilient Big Data Monetization* is characterized by being decomposed into two subproblems, namely *k*-dominance (Theorem. 3) and *m*-connectivity (Theorem. 4), such that common-interests are connected to *k* measurement tools, who are tied within each other in *m*-ways.

## C. Proposed Scheme

In this context, a greedy approximation algorithm *Plutus* (Fig. 2(c)) is proposed, which assists in acquiring sustainable domination of measurement tools over common-interests. In this process, influential measurement tools are isolated (Isolation) to acquire dominance over common-interests (Domination), which are gradually dominated by multiple measurement tools (Synergie), who are tied each other (Divergence) and gain sustainability (Sustainability). Hence, *Plutus* resolves k-dominance and m-connectivity problems with Isolation, Domination, Synergy, Divergence and Sustainability models, respectively.

## D. Theoretical Analysis

Our theoretical analysis is summarized as follows

(a) *Resilient Big Data Monetization* is a NP-hard problem, which is amenable to greedy approximation.

(b)*Plutus* lays out the theoretical foundation of *Resilient fact-finding*(i.e inferring resilience (i.e fact) from the observed information), which is characterized by being decomposed into four major properties

Definition 1. *Maximal Independent Fact-Finding: Maximal Independent Fact-Finding is aimed at inferring an independent set(i.e fact) M(G) V (G), such that 8(u; v) in G, no edge exists between uand v and no vϵ(G M) can be added to M. Hence, this is resembled by isolation of dominant measurement tools, which assist service provider to monetize from Big Data.*

Definition 2. *Influential Fact-Finding Influential Fact-Finding is aimed at inferring influential set (i.e fact) D(G)of G, if 8vϵG either vϵD(G) or 9u such that(u; v)ϵE(G) and graph induced byD(G) is connected. Hence, this corresponds to acquiring influence of measurement tools over common interests, which depict personalized services for Smart-device users.*

Definition 3. *k-dominance Fact-Finding: k-dominance Fact-Finding is aimed at inferring a dominating set (i.e fact) D(G)ϵV (G) , such that 8vϵV (G) D(G), v is adjacent to at least k nodes in D(G). Hence, this is significant with ties between common-interest and measurement tools, which assist common-interest in achieving synergic domination by multiple measurement tools.*

| Big Data Monetization | Measurement-tool | Common-interest |
| --- | --- | --- |
| Connected Dominating Set | Dominator | Dominate |
| Fact-Finding | Observation(*x*) | Parameter ( ) |

TABLE I
MAPPING TERMINOLOGIES OF *Big Data Monetization*, CONNECTED DOMINATING SET, FACT-FINDING

Definition 4. *m-connected Fact-Finding: m-connective Fact-finding is aimed at inferring a dominating set (i.e fact)D(G)ϵV (G) , such that graph induced by D is m-connected, representing D is connected after m-1 dominators are removed. Hence, this is represented by ties among mea-surement tools, which assist measurement tools in acquiring divergent and sustainable domination over common-interest.*

## E. Organization

This paper is organized as follows, (a)*Resilient Big Data Monetization* problem is formulated in Section II, (b) *Plutus* is proposed in Section III, (c) Theoretical Analysis is presented on Section IV (e) Related works and Conclusion are presented in section V and VI, respectively.

## II. PROBLEM FORMULATION

In this section, *Resilient Big Data Monetization* problem is formulated with preliminary definitions of measurement, inference, Big Data monetization.

Let *v* and *u* be measurement tool and common-interest, respectively in Big Data Monetization scenario. Let us assume that Big Data Monetization is aimed at deploying measurement-tool*v* is inferring common-interest *u* by observ-ing its usage-dynamics.

Definition 5. *Measurement is defined as the procedure through which v quantifies usage-dynamics of u , prior or upon or after incentivization.*

Definition 6. *Inference is defined as the procedure through which the u is predicted/depicted by v.*

Definition 7. *Big Data Monetization is defined as the incentivizaton procedure, which facilitates v to depict u after quantifying usage-dynamics.*

Lemma 1. *Big Data Monetization is decomposed into two sub-problems, namely (a) Measurement and (b) Inference. [Definition [1],[2],[3]].*

Theorem 1. *Big Data Monetization is a dominating set problem.*

*Proof.* Big Data monetization is aimed at facilitating dominance of measurement tools over common-interests. Given *v; u*, it is to derive a graph *G*, so that for every *v* belongs to *G*, either *v* belongs to *D(G)* or there exists a *u* such that (*u; v*) belongs to *E(G)*. Hence, Big Data monetization is a dominating set problem. □

Theorem 2. *Resilient Big Data Monetization is a connected dominating set problem.*

*Proof.* The domination of measurement tools over common-interests is best-achieved by the ties not only in-between measurements, but also between measurements and common interests. Given $v$ and $u$, it is to to derive $D(G)$ $V(G)$, which is a dominating set of $G$ and graph induced by $D(G)$ is connected. Hence, Big Data monetization is a connected dominating set problem. □

Theorem 3. *Measurement is a m-connectivity problem.*

*Proof.* The quantification of usage-dynamics is best-achieved with measurement tools, who are tied to each other. Given $v$ and $u$, it is to derive a connected dominating set $D$, which is connected after $m$ 1 alternations of $v$. Hence, resulting dominating set $D(G)$ $V(G)$ is $m$-connected. Hence, Measurement is a $m$-connectivity problem. □

Theorem 4. *Inference is k-dominance problem.*

*Proof.* Inference necessitates that each common-interest is quantified through diversified perspectives from multiple measurement tools. Given $v$ and $u$, $u$ is adjacent to at least $k$ $v$ in $D(G)$. Hence. resulting graph is $k$-dominating set $D(G)$ $V(G)$. Therefore, Inference is a $k$-dominance problem. □

Definition 8. *Resilient Big Data Monetization is resembeled in between v and u, such that m-alternations of v assist in acquiring k-dominance on u.*

Lemma 2. *Resilient Big Data Monetization is a m-connectivity k-dominance problem (Theorem [3],[4], Defini-tion [4])*

### III. PROPOSED SCHEME

In this section, a greedy algorithm *P lutus* (Algorithm 1) is proposed, which assists in obtaining sustainable dominance of dominators over dominate. In this process, at first, dominators are isolated (Isolation), then simple domination is obtained over dominate (Domination), which gradually reaches synergic stage (Synergy). Finally, dominators earn diversified domina-tion (Diversity), which gradually reaches at sustainable stage (Sustainability). Hence, *P lutus* resolves $k$-dominance and $m$-connectivity problems with Isolation, Domination, Synergy, Diversification and Sustainability, respectively.

---
Algorithm 1: Plutus

Data: All nodes
Result: D(3-connected k-dominating set)
Round 1:Isolation (Algorithm 2)
Round 2:Domination (Algorithm 3)
Round 3:Synergy(Algorithm 4)
Round 4:Divergence (Algorithm 5)
Round 5:Sustainability (Algorithm 6)

---

In Isolation (Algorithm 2), all nodes are regarded as domination-prone (Fallacy) and then gradually converted to either dominator or domination-reluctant (Separation). At first (Fallacy), all nodes are initially assigned as domination-prone.

---
Algorithm 2: Isolation

Data: All nodes
Result: dominator and domination-reluctant node (Fallacy)
-Initialize all nodes as domination-prone nodes -Choose node with maximum cardinality as dominator -Assign neighbor and next-to-neighbor as domination-reluctant and domination-prone nodes, respectively.
(Separation)
while *domination prone node exists* do
 -Find domination-prone node with the highest domination-reluctant neighbor;
 -Assign dominator role to it
 -Assign its neighbor as domination-reluctant. end

---

However, node with the highest cardinality is assumed as dominator, its neighbors and next-to-neighbors are regarded as domination-reluctant and domination-prone, respectively. Then, the most domination-prone (i.e. node having highest domination-reluctant neighbors) is chosen and assigned the dominator role. Then, assign neighbor of newly created dom-inator as domination-reluctant. The process continues until domination-prone disappears (Separation). Hence, Isolation ends up with dominator and domination-reluctant.

---
Algorithm 3: Domination

Data: dominator and domination-reluctant
Result: D(1-connected 1-dominating set)
forall the *Dominator Pair* do
 (Trimming)
 -Compute shortest path between every pair of dominator
 (Convergence)
 -Convert intermediate dominator-reluctant as dominator
 -Assign dominator pair and newly created dominator (i.e domination-prone) to D
 -Assign neighbor and next-to-neighbor as domination-reluctant and domination-prone, respectively.
end

---

In Domination (Algorithm 3), domination-reluctant are gradually converted to dominator. In this process, firstly short-est path between every pair of dominator (u, v) is calculated (Trimming). Then, intermediate domination-reluctant of every pair of dominator (u,v) are considered as dominator (Con-vergence). Therefore, Domination generates 1-connected 1-dominating set at the end.

In Synergy (Algorithm 4), k-domination over dominate is acquired by dominator. In this process, firstly isolated dominators (i.e output of Isolation (Algorithm 2)) are sepa-

### Algorithm 4: Synergy

Data: D(1-connected k-dominating
set) Result: D(2-connected 1-
dominating set) (Displacement)
-Remove isolated dominator (output of Algorithm
1) from the graph
(Adjustment)
for *i=2 to K* do
> -Isolate dominator set in G-$M_1$ $M_2$...$M_{i\,1}$,
> consecutively
> -Add to D
end

---

rated from graph (Displacement). Then, dominators are again isolated from resulting graph and added to D (Adjustment). The adjustment process continues k-times until k-divergence is achieved.

### Algorithm 5: Diversification

Data: D(1-connected k-dominating set)
Result: D(2-connected k-dominating set)
-Find all blocks in 1-connected k-dominating set
while *dominators are not 2-connected do* do
> (Blocking)
> Derive blocks in graph
> (2-Connectivity)
> Add all intermediate nodes of shortest path that
> connects leaf block in D to other part of D, so that it
> does not have any nodes in D except two endpoints
end

---

In Diversification (Algorithm 5), D acquires 2-connectivity by facilitating connectivity to leaf blocks. In this process, all blocks are derived (Blocking) and then all intermediate nodes of shortest path are connected in a way, such that leaf block in D is connected to other part of D and it does not have any nodes in D except two endpoints (2-Connectivity). The iteration goes on until 2-connectivity is achieved.

### Algorithm 6: Sustainability

Data: D(2-connected k-dominating set)
Result: D(3-connected k-dominating
set) while *There is no bad-point do* do
> -Convert bad point to good point by moving from
> G-D to D
end

---

In Sustainability (Algorithm 6), 3-connectivity is acquired by dominator by bad-point removal. In this process, bad points are iteratively converted to good point by moving from *G-D* to *D*, such that no new bad point is created. The process continues until bad point disappears. In this way, finally 2-connected *k*-dominating set acquires 3-connectivity.

## IV. THEORETICAL ANALYSIS

In this section, *Plutus* is justified to lay out the theoretical foundation of Resilient Fact-Finding, resembled by maximal independent, domination, *k*-dominance, 2-connectivity and 3-connectivity fact-finding. It is followed by the justification how Resilient Big Data Monetization is a NP-Hard problem and amenable to greedy approximation.

**Theorem 5.** *Isolation lays out the theoretical foundation of Maximal Independent Fact-finding*

*Proof.* In Isolation (Algorithm 2), when a node is extracted as dominator, its neighbor and next-to-neighbor are regarded as domination-reluctant and domination-prone, respectively. However, only the most domination-prone node (i. e having highest domination reluntant node) joins isolated dominator set at the next round. The process continues until there is no domination prone node exists. In other sense, the process ends up with isolated dominator set, accompanied by domination reluctant nodes. So, there is no possibility of inclusion of neighbors in isolated dominator set.

Let $v; u; u'$ be dominator, domination-reluctant, domination-prone nodes, respectively.

Therefore, at initialization $v$ joins $I(G)$. Then, $u' \epsilon G$ is turned to $v$ by joining $I(G)$; such that $\partial w : w \epsilon u, (w; u') \epsilon E(G)$ appears as the most frequent,

$I(G)$ $V(G)$ is an independent set of G, since $\partial(u; v)$ no edge exists between $u$ and $v$

The procces finishes, when $u'$ does not exist. Hence inde-pendent set $I(G)$ is a maximal independent set $M$, since no $v \epsilon (G\ M)$ can be added to $M$. If any $v \epsilon (G\ M)$ is added, it is not an independent set anymore.

Hence, Isolation lays out the theoretical foundation of Maximal Independent fact-finding.
□

**Theorem 6.** *Influence lays out the theoretical foundation of Influential Fact-finding.*

*Proof.* In Influence (Algorithm 3), the shortest path between every pair of isolated dominators are measured and domination-reluctant nodes found on the route are assigned to CDS.

Let, $G; I; D$ be connected graph, independent set and dominating set. Assume, for any pair of nodes $u$, $v$ with $(u; v) = 2; dD(u; v) <= \_ + 1$. Therefore, for any pairs of distinct nodes $u$ and $v$, $d_D(u; v) <= \_ d(u; v)$

Assume, $I$ is a subset of $D$ such that for any pair of vertices $u; v$ in $I$ with $d(u; v) < 4$, $dD <= 4$ So, for every pair of distinct nodes $u; v$ $d_D(u; v) <= 5d(u; v)$

All intermediate domination-reluctant nodes in the shortest path of two dominatoor nodes $u; v$ (where $d(u; v)$ 4 ) are turned to dominator. Hence, intermediate domination-reluctant nodes and dominators construct connected dominating set. Hence, Influence lays out the theoretical foundation of influ-ential fact-finding.
□

**Theorem 7.** *Synergy lays out the theoretical foundation of k-dominance fact-finding.*

*Proof.* Synergy intends on deriving *k*-way connections from dominate to dominator.

Let, *G; D; I* be connected graph, connected dominating set and maximal independent set respectively.

First, isolated dominators (*I*) are extracted and then k-1 subsequent isolated dominators are added back to *D*.

Hence, each dominate (*G D*) are k dominated by dominator(*D*). That represents that each dominate is connected to at least *k* dominator. Therefore, Synergy lays out the theoretical foundation of *k*-dominance fact-finding. □

**Theorem 8.** *Diversification lays out the theoretical foundation of* 2-*connectivity fact-finding.*

*Proof.* Diversification aims that dominators are resilient upto (*m* = 2) 1 alternations. Hence, desired solution is intended on moving all dominators in the same block.

Let *v∈D*(*G*) be a cut vertex, such that *D v* ends up with disconnected dominating set. Hence, a block is required, which does not include any cut-vertex.

In this context, a leaf block (i.e. a block, having one cut-vertex ) is discovered and connected to other part of *D* in a ways, so all dominators are in a maximal connected subgraph of *D*, denoted as block.

Therefore, dominators are 2-connected, as sub-graphs have no cut-vertex, the disjunction of which might leave graph disconnected. Hence, diversificatin lays out the theoretical foundation of 2 connectivity fact-finding. □

**Theorem 9.** *Sustainability lays out the theoretical foundation of* 3-*connectivity fact-finding*

*Proof.* Sustainability aims that dominators are connected after (m=3)-1alternations.

Let us assume that *v∈D*(*G*) be good or bad point, if subgraph induced by *D fvg* is still 2-connected or not, respectively.

Hence, discovering a bad point in 2-connected graph and then removing it our desired solution. Since, 2-connected graph without bad-point is considered 3-connected. In this context, bad point is converted to good point by moving into dominating set.

3-connectivity is resembled by removing at least three nodes to disconnect *D*. Therefore, if *v* be a good point in 2-connected graph *D*, *D fvg* is 2-connected. In other sense, dominators are resilient upto 2 more altenations of connections between them. Hence, sustainability lays out the theoretical foundation of 3-connectivity fact-finding. □

**Lemma 3.** *Resilient Big Data Monetization is a NP-hard problem.*

*Proof.* CDS-construction is a NP-hard problem in UDG. Hence, *m*-connected *k*-dominating set with uncertainty con-straint, and thereby devised problem is a NP-hard prob-lem. □

**Lemma 4.** *Resilient Big Data Monetization is amenable to Greedy approximation*

*Proof.* No PTAS exists for weighted CDS construction. Hence, no PTAS exists for *m*-connected *k*-dominating set an uncertainty contraint and thereby for devised Problem. Hence, devised problem is amenable to greedy approximation. □

## V. RELATED WORK

In this section, state-of-the art Big Data management mon-etization, measurement, inference, connected dominating set, fact-finding are presented.

### A. Big Data Monetizatin Schemes

Social reputation based monetization scheme[27] measures high quality participatory information by excluding those, which do not comply with prescribed social strategy. How-ever, this scheme is resilient/adaptive to time-dependency of current and future behavior of Smart-device user. An auction based monetization scheme[5] also measures truthful cost-reporting through a participation-level payment scheme. However, this scheme merely utilizes private participation information with quality-of-service, rather than common-interests. Similarly, Game-theoretic monetization scheme[28] measures lowest-price reserved by Smart-device user not to pay less-amount in return on auction. However, this scheme is resilient to interest of those Smart-device users, who only tailor their actions to cater for the platform. Heuristic mechanism[29] measures sensory data to be offloaded via 3G. However, the scheme considers the interest of Smart-device users, who seek availability of WiFi at access-points. Another heuristic monetization scheme[30] measures location-traces from Smart-devices. However, the scheme is motivates users based on their common interest, such as location or sometimes emotion. Psychological-trait based monetization[31] measures psychological insights of human (e.g affection-to or accessibil-ity/price of incentive). However, their scheme is intended to a specific interest (i.e prefetching media content) of user-group, who come within WiFi hotspot. Monetization scheme[32] often measures both individual and social welfare of Smart-device participant. Hence, the scheme is intended on user-groups having not only service-consumption demand, but also an optimal participation level. Auction based monetization mechanism[33] measures Smart-device users bid prices and dropping rate. However, the scheme is dedicated to dy-namic true valuation of Smart-device users, sensing-data or event context. However, micro-payment-based monetization scheme[34] measures dynamic monetary support for Smart-device users with the advent of temporal and spatial coverage. However, the scheme cares about the user-group, who carry altruistic and competitiveness at participation. A heuristic monetization[13] also measures both social reputation and participation-interest of Smart-device users. However, the scheme considers interest-genre of user, who might be either motivated or reluctant about participation.

*B. Challenges of Resilient Big Data Monetization*

Apart from usage-duration, real-time feedback, or renovated tools[1][6][7], often keyword extraction scheme is trained with large corpus of data[14]. Moreover, psychological states, such as expectations and purposes of users often come useful as measurement tool[17].

Service provider continuously strives for achieving domination of measurement tools over targeted common-interests[15][16]. Hence, they seem to classify measurement tools to infer social contexts and update their classification mechanism from usage-history[15]. Often measurement tools are ranked, weighted by people's expertise[16].

Discovering binding between common-interests and measurement tools is of significance. Hence, correlating activity-context (e.g location) and quantifying ties between environmental characteristics (e.g air and weather) and pollutant concentration are frequently observed nowadays[18]. Smart-phone traces, real-time feedback are often connected with strength of sociability or relations with colleagues[19]. Hence, often incentivization mechanisms are designed to utilize correlations between social behavior and usage-frequency[20].

Similarly, grouping or hierarchy among measurement tools becomes useful, especially when observed information is diversified, such as fingerprint, ambient, sound, light or color. On some occasions, grouped observations give better measurement results. For example, Smart-phone traces from waiting people and on-bus people are grouped to infer bus-route or bus-arrival time[21]. Similarly, well-being monitoring is quan-tified better through grouped observations from different time of a day, diversity of network, entropy of movement within or outside or even diversity of face-to-face interactions[22], etc.

Once measurement-tools are tied up, it is necessary that it sustains. Hence, connections between Smart-phone GPS-trace, feedbacks, speaking, verbal and even proximity interaction of social users are often grouped with statistical learning tools[23][24]. Sometimes, Smart-phone ambient information, such as acoustic (e.g. ringtone and acoustic sound) and visual (e.g light intensity and view similarity) signs are ranked through grouping[25]. Even, Smart-phone, sensor and social network traces are grouped to infer new pattern, especially in social-behavior measurement[26].

*C. Measurement*

Conventional measurement tools, such as download-count, user-rating, usage-duration are extensively utilized by service providers to infer usage-dynamics of customers[1]. However, the recent proliferation of Smart-devices and relevant person-alized services motivate devising new measurement tools or techniques[6][7].

In this context, users feedback, service-usage information, traffic pattern are measured to discover both qualitative variance and quantitative similarities of Smart-device users. Application or service-specific (e.g advertisement, media-traffic) measurement tools are devised as well. Advertisement-playing or even user-retention time or even Ad-skipping tendencies are common measurement tools in Smart-Ad management. On the other hand, buffering time, video-quality and introductory la-tencies are regarded as emerging media-content measurement tools[7].

The penetration of Smart-devices also has ended up with proliferation of Smart-device apps, which measure user-engagement in innovative ways. Users engagement is measured to predict his/her usage-duration and longevity/affection on a specific Smart-device app. Often, collaborative measurement technique is devised with usage-duration, usage-dynamics and real-time feedback. Moreover, users reaction, such as head-movement, talking are often regarded as quantitative engagement measurement tools[1][6].

*D. Inference*

Emotion, location or environmental information inferred from Smart-device traces are frequently utilized in personal-ized services.

Smart-device users emotional or behavioral pattern is of-ten inferred through obtrusive or unobtrusive way to pro-vide personalized shopping, search or health-care support. Emotion inferred from mobility, weather, time and activity assist in customized search experience for user[2]. However, behavior or movement pattern plays a vital role at well-being monitoring[8]. Moreover, group behavior of users in movement in a shopping mall assist in personalized product recommendation[9].

On the other hand, users location often assists in traf-fic signal detection, transit-tracking, sociability or consumer-applications. GPS location traces are often utilized to recommend cheaper grocery product in nearby shops[3]. Often WiFi-tracks are utilized to predict optimal user-experience in wireless connectivity[10]. Collaboration among multiple traces often assist in tracking traffic penetration in a vicinity or even monitoring movement of transit[3].

Similarly, traces from Smart-device users are utilized to infer environmental characteristics (e.g. local weather, air-pollution)[11][12], earthquake or even nuclear threat in a city[4].

*E. Connected Dominating Set*

State-of-the-art CDS research is categorized based on net-work model (i.e graph, UDG or UBG) or resilience (i.e m-connectivity and/or k-dominance).

CDS is justified to lack PTAS scheme in general graph model[35]. However, in UDG, even though an approxima-tion solution is discovered, it is beyond expected optimal solution[36]. Since then devising faster approximate solution has been a major concern in CDS research. Performance ratio approximation is based on conventional two-phases of CDS construction, namely MIS and CDS construction. Hence, one common trend is to approximate optimal solution by observing MIS with minimum CDS size[36]. Accordingly, the performance ratio is approximated in terms of MIS size.

However, another trend approximates performance ratio from CDS construction with uncertainty constraints, which is char-acterized by routing cost, load-balance or even diameter[37].

Meanwhile, the intrinsic uncertainty involved in node or edge-connectivity inspires the inclusion of resilience in CDS construction[38][39][40]. However, as even the simplest case (m=1, k=1) of m-connected k-dominating set is NP-hard, worst case performance guarantee comes as the first design choice. Hence, greedy approximation algorithm is proposed to con-struct 1-connected 1-dominating set , then make it 2-connected and augment subgraph by adding path iteratively. Minimum weight Steiner network is devised to satisfy connectivity requirement with the inclusion of new nodes. However, our proposal is merely intended on utilizing existing nodes, rather than including new one. General Steiner network is devised to extract a subset of nodes to facilitate edge-connectivity. Similarly, Spanning tree is often deployed to extract minimum weight or edge-connectivity. In contrast, our proposal is aimed at isolating or then connecting nodes. Gossip-based or graph coloring and coverage based deterministic approaches are de-vised to acquire only k-dominations on dominate, which often lack upper bound on CDS size. However, our approach not only facilitates k-domination, but also acquires m-connectivity with approximation ratio.

### F. Fact-Finding

Fact-finding[41] is devised to iteratively estimate the cred-ibility of source and claim to end up with fact-inference from observed information. Voting[42], an earlier version, is believing on the fact, which is believed by majority of sources. Page-rank[43] and truth-finding[44] are aimed at iteratively inferring the credibility of source and claim from the simple relationship between claims. Bayesian[45], max-imum likelihood[46] or Cramer-Rao bound[47] based fact-finding mechanism are devised to infer the fact from statistical learning information. Moreover, the hardness of claims are often analyzed through cosine, 2-estimate and 3-estimate[48]. However, most of the schemes are based on the assumptions on independent facts. In contrast, our devised fact-finding mechanisms are aimed at inferring the grouping behavior in between sources(i.e dominator) and also between source(i.e dominator), claim(i.e dominate).

## VI. CONCLUSION

The proliferation of Smart-device enabled personalized services motivates designing new Big Data monetization schemes. In this context, a greedy approximation solution *Plu-tus* is proposed, which assists in acquiring dominance of mea-surement tools over common-interests of Smart-device users. *Plutus* lays out the theoretical foundation of *Resilient Fact-Finding*, resembled by isolation of dominant measurement tools over common-interest, followed by binding common-interests to measurement tools, which are tied within each other to prevail.